\documentclass[10pt, conference, compsocconf]{IEEEtran}
\usepackage{amssymb,amsmath,epsfig}
\usepackage{stmaryrd}
\usepackage{cite}
\usepackage{algorithmic}
\usepackage{algorithm}
\usepackage{graphicx}
\usepackage{amssymb}
\usepackage{amsthm}
\usepackage[keeplastbox]{flushend}
\begin{document}
\title{Identification of GSM and LTE Signals Using Their Second-order Cyclostationarity}

\author{\IEEEauthorblockN{Ebrahim~Karami\IEEEauthorrefmark{1},~Octavia~A.~Dobre\IEEEauthorrefmark{1}, and~Nikhil~Adnani\IEEEauthorrefmark{2}}\\
\IEEEauthorblockA{\IEEEauthorrefmark{1}Electrical and Computer Engineering, Memorial University, Canada}
\IEEEauthorblockA{email: \{ekarami,odobre\}@mun.ca}\\	
\IEEEauthorblockA{\IEEEauthorrefmark{2} ThinkRF Corp., Ottawa, Canada}
\IEEEauthorblockA{email: info@thinkrf.com}}

\maketitle
\begin{abstract}
Automatic signal identification (ASI) has various millitary and commercial applications, such as spectrum surveillance and cognitive radio. In this paper, a novel ASI algorithm is proposed for the identification of GSM and LTE signals, which is based on the pilot-induced second-order cyclostationarity. The proposed algorithm provides a very good performance at low signal-to-noise ratios and short observation times, with no need for channel estimation, and timing and frequency synchronization. Simulations and off-the-air signals acquired with the ThinkRF WSA4000 receiver are used to confirm the findings.

\end{abstract}

\let\thefootnote\relax\footnotetext{This work was supported in part by the National Sciences and Engineering Research Council of Canada (NSERC) through the Engage program.}

\section{Introduction}
Automatic signal identification (ASI) has been initially investigated for military communications, e.g., for electronic warfare and spectrum surveillance \cite{Dobre1}. More recently, ASI has found applications to commercial communications, in the context of software defined and cognitive radios \cite{j_Mitola,yucek2009survey}.\par
ASI tackles the problem of identifying the signal type without relying on pre-processing, such as channel estimation, and timing and frequency synchronization \cite{Dobre1,yucek2009survey,Kim,Adrat,Bouzegzi,Ciblat1,Alaa1,
Alaa2,Alaa3,Alaa4,gardner1992signal,Dobre_MOD_Cyclostationarity,classf_m2}. While most of the ASI work in the literature has been done for generic signals, very few papers investigate the identification of standard signals; however, the latter is crucial for spectrum surveillance and cognitive radio applications. ASI techniques usually exploit signal features to identify the signal type \cite{Kim,Adrat,Bouzegzi,Ciblat1,Alaa1,Alaa2,Alaa3,Alaa4,gardner1992signal,
Dobre1,Dobre_MOD_Cyclostationarity}, and the feature-based identification of standard signals has been carried out as follows. In \cite{Kim}, the authors use second-order cyclostationarity-based features to classify different IEEE 802.11 standard signals. The pilot-induced cyclostationarity of the IEEE 802.11a standard signals is studied in \cite{Adrat}, with ASI application. Kurtosis-based features are proposed in \cite{Bouzegzi,Ciblat1} to identify OFDM-based standard signals. Furthermore, the cyclic prefix (CP)-, preamble-, and reference-signal-induced second-order cyclostationarity of LTE and WiMAX standard signals is exploited in \cite{Alaa1,Alaa2,Alaa3,Alaa4} for their identification. While the previously mentioned feature-based ASI techniques are developed for orthogonal frequency division multiplexing (OFDM)-based signals, they are not necessarily appropriate to other standard signals, such as GSM. Therefore, to identify such standard cellular signals, we need to develop ASI algorithms based on new features.  
In wireless communications systems, pilot signals are used for channel estimation, as well as frequency and timing synchronization. As the pilot symbols are sent periodically, one can use this periodicity to identify different wireless standard signals.
In this paper, we propose a low complexity algorithm to identify the GSM and LTE standard signals, as being widely used in Canada; off-the-air signals are used for verification. \par
The rest of the paper is organized as follows. Section \ref{sec:System-Model}
presents the model for the GSM and LTE standard signals. Section \ref{sec:Alg} introduces the proposed algorithm for the identification of these signals. In Section \ref{sec:Results}, results for off-the-air signals acquired with the ThinkRF WSA4000 receiver are shown, along with simulation results. The paper is concluded in Section \ref{sec:Conclusion}.

\section{Signal Model\label{sec:System-Model}}
In this section, the signal model for the GSM and LTE downlink (DL) is introduced. More specifically, we present the pilot signals in these standards, as their periodicity will be exploited for the identification feature. 

\begin{figure}
\centering
\includegraphics[width=1\linewidth]{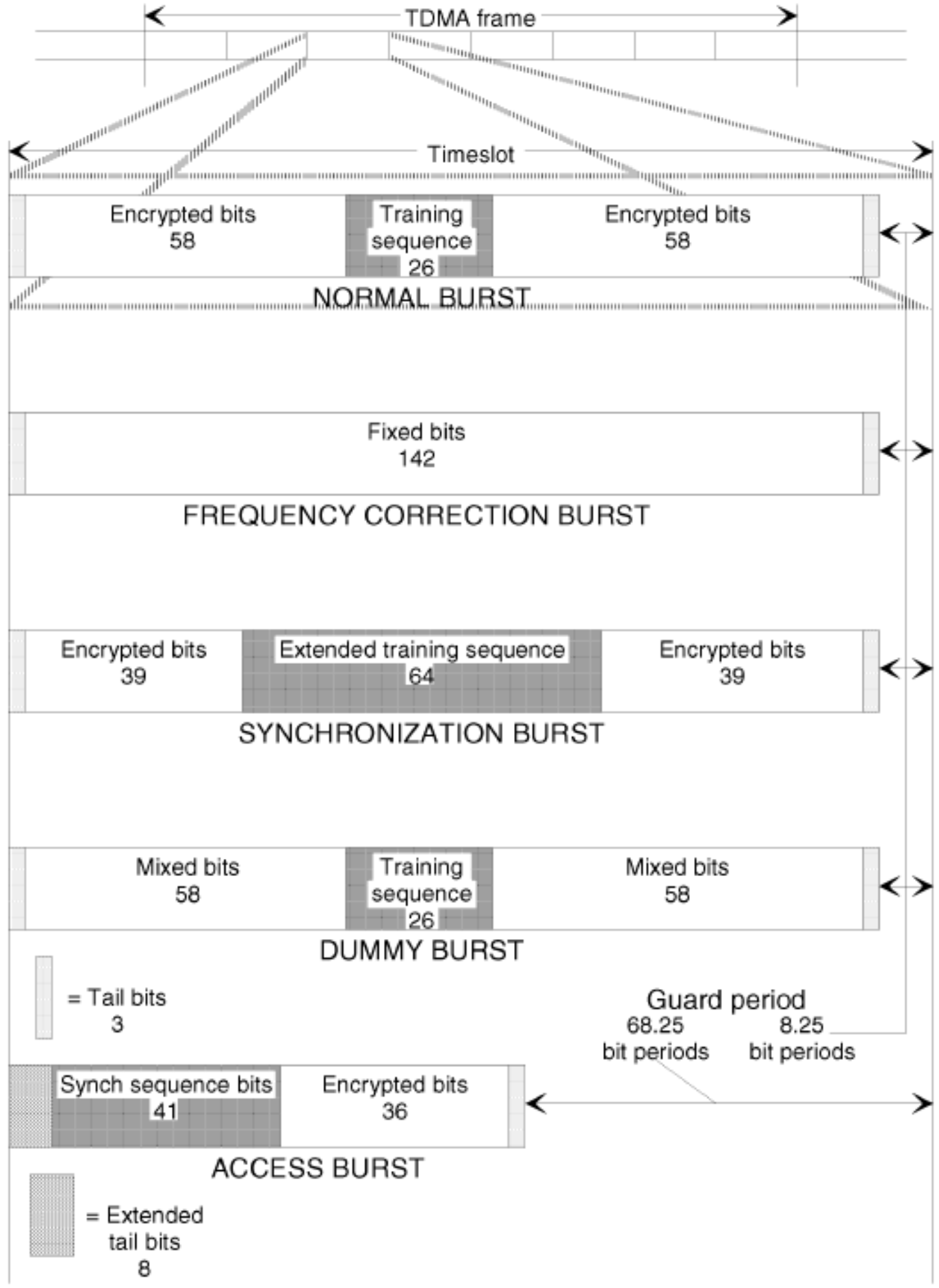}
\caption{Time slot and format of bursts in the GSM systems\cite{ETSI0502}.}
\label{GSM_frame}
\vspace{-1.2em}
\end{figure}

 \begin{flushleft}
 \textbf{A. GSM Signal Model}
 \end{flushleft}
The GSM frame structure is shown in Fig. \ref{GSM_frame}, including the normal burst, which carries data, the control bursts, such as frequency correction and synchronization, as well as the access bursts \cite{ETSI0502}. In the normal burst,  26 bits in each time slot are dedicated  to training; these are repeated every time slot and used for channel estimation. Since the duration of each time slot is 577 $\mu$s, the repetition frequency of the pilot sequence is 1733 Hz. From Fig. \ref{GSM_frame}, one can see that the other GSM bursts have similar repetitive sequences, but with different lengths; however, all repeat with the same frequency, i.e., 1733 Hz.

\begin{figure}
\centering
\includegraphics[width=1\linewidth]{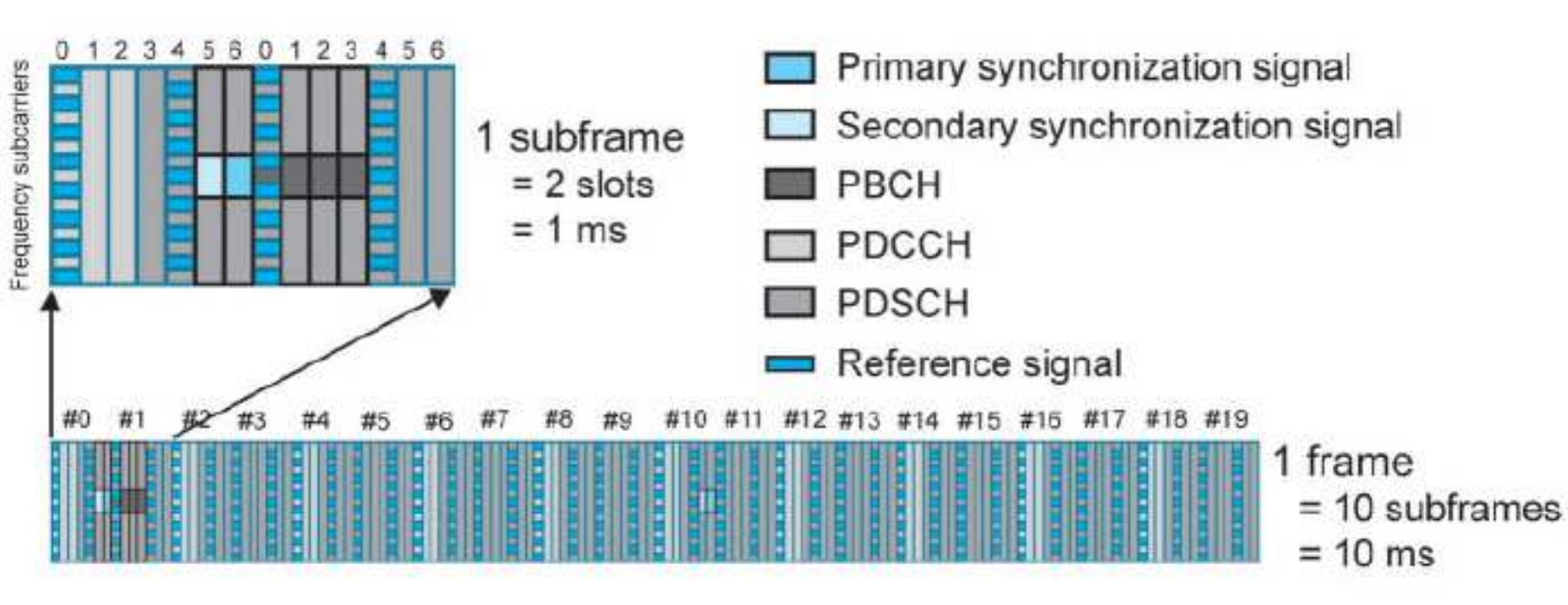}
\caption{LTE FDD DL frame structure\cite{book_lte0}.}
\label{LTE_frame}
\vspace{-1.2em}
\end{figure}

 \begin{flushleft}
 \textbf{B. LTE DL Signal Model}
 \end{flushleft}
The LTE frequency division duplex (FDD) DL frame structure is shown in Fig. \ref{LTE_frame}. Each LTE frame includes 20 time slots, each with 6 or 7 OFDM symbols, depending if the short or long CP is used \cite{book_lte}. In Canada, LTE with short CP is commonly employed. From Fig. \ref{LTE_frame}, one can see that the samples which are periodically repeated correspond to the cell specific reference signals (RSs), and primary and secondary synchronization channels (PSCH and SSCH), where the RS is repeated every time slot and PSCH and SSCH are repeated every 10 time slots. The duration of each LTE time slot is 0.5 ms. Consequently, the repetition frequency for the RSs is 2 kHz, while for the PSCH and SSCH is 200 Hz.

\section{Proposed Signal Identification\\Algorithm \label{sec:Alg}}
In this section, the algorithm proposed for the identification of GSM and LTE standard signals is presented. First, we introduce the fundamental concept of signal cyclostationarity in order to further discuss the identification feature, and then present the feature-based algorithm and study its complexity.

\subsection{Second-order Signal Cyclostationarity}
A signal $r(t)$ exhibits second-order cyclostationarity if its first and second-order time-varying correlation functions are periodic in time \cite{gardner1994cumulant}. In this work, the following second-order time-varying correlation function is considered
\begin{equation}\label{eq:corr}
 c(t,\tau)=\mathrm{E}\left[r(t)r^*(t+\tau)\right],
\end{equation}

\noindent where $*$ denotes complex conjugation, $\mathrm{E}\left[.\right]$ is the statistical expectation, and $\tau$ is the delay. If $c(t,\tau)$ is periodic in time with the fundamental period $M_0$, then it can be expressed
by a Fourier series as \cite{gardner1994cumulant}
\begin{equation}
c(t,\tau)=\sum_{\{\alpha\}}C(\alpha,\tau)e^{j2\pi t \alpha}.\label{eq:Fourier_series}\end{equation}

 The Fourier coefficients defined as 
 \begin{equation}
C(\alpha,\tau)=\frac{1}{M_0}\int_{0}^{M_0}\!c(t,\tau)e^{-j2\pi t \alpha}\,\mathrm{d}t,\label{eq:CCF}
 \end{equation}
\noindent are referred to as the cyclic correlation function (CCF) at cyclic frequency (CF) $\alpha$ and delay $\tau$.  
The set of CFs is given by $\{\alpha\}=\{\frac{\ell}{M_0}, \, \ell \in \mathbb{\mathcal{\mathcal{I}}} \, \mbox{, with} \, \mathbb{\mathcal{\mathcal{I}}} \, \mbox{as the set of integers}\}$.
Assuming $M_r$ as the number of received samples, CCF at CF $\alpha$ and delay $\tau$ is estimated from the received sequence, $r(m)$, as \cite{Test_cyc1}
\begin{equation}
\hat{C}(\alpha,\tau)=\frac{1}{M_r}\sum\limits_{m=0}^{M_r-1}r(m)r^*(m+\frac{\tau}{T_s})e^{-j2\pi\alpha m T_s}. \label{est_CCF}
\end{equation}
\noindent where $T_s$ is the sampling period and $\tau$ is multiple integer of $T_s$. \par
Due to the periodicity of the pilot signals in GSM and LTE standards, one can show that these induce second-order cyclostationarity with CFs $\alpha_i=\frac{\ell}{T_i}$, $i=$GSM, LTE, where $T_i$ is the time slot duration of the GSM and LTE standards. The pilot-induced second-order cyclostationarity will be used as an identification feature, as presented in the next sub-section.
\subsection{Proposed Second-order Cyclostationarity-based Algorithm}
We explore the CCF at CF $\alpha$ and zero delay $C(\alpha,0)$ to identify the GSM and LTE standard signals, as follows. In the first step, $\hat{C}(\alpha,0)$ is estimated at CFs $\alpha_i=\frac{1}{T_{i}}$, $i=$GSM, LTE. 
In the second step, the estimated CCF magnitude is compared with a threshold, which is set up based on  a constant false alarm criterion. The probability of false alarm is defined as the probability of deciding that the standard signal is present when this is not (either an unknown signal or noise is present). An analytical closed form expression of the false alarm probability is obtained based on the distribution of the CCF magnitude estimate for the unknown signal and noise; in this case, one can simply infer that the CCF magnitude estimate has an asymptotic Rayleigh distribution \cite{Test_cyc1}. Hence, if the CCF for a specific CF $\alpha$ and delay $\tau$ is used as a discriminating feature, the probability of false alarm is calculated using the complementary cumulative density function of the Rayleigh distribution as 
  \begin{equation}
P_F=\exp(-\frac{\Gamma^2}{\sigma_r^2}), \label{eq:PF}
\end{equation}
\noindent where $\sigma_r^2$ is the variance of the received signal. A summary of the proposed algorithm is provided as follows.

\floatname{algorithm}{}

\begin{algorithm}
\renewcommand{\thealgorithm}{}
\caption{\textbf{Proposed algorithm}}
\begin{algorithmic}[0]
\STATE \textbf{Input:} The received sequence $r(m)$, $m=0,...,M_r-1$.
\STATE - Estimate the CCF, $C_i=\hat{C}(\alpha_i,0)$, using (\ref{est_CCF}) at CFs $\alpha_i=\frac{1}{T_{i}}$, $i=$GSM, LTE.
\STATE - Estimate the variance of the received signal, $\sigma_r^2$, and calculate the threshold using (\ref{eq:PF}).
\IF{$C_{i}> \Gamma$}
\STATE - {The received signal is identified as $i$, $i=$GSM, LTE.}

\ELSE
\STATE - {The type of the received signal is not $i$ and it can be either an intruder or noise.}
\ENDIF
\end{algorithmic}
\end{algorithm}
\par
\textit{Computational complexity:} We evaluate the computational complexity of the proposed identification algorithm through the number of floating point operations (flops) \cite{matrixcomp}, where a complex multiplication and addition require six and two flops, respectively. Based on (\ref{est_CCF}), one can easily see that the number of complex multiplications and additions needed to calculate the CCF equals $2M_r$ and $M_r-1$, respectively. By considering that the thresholding does not require additional complexity, it is straightforward that the number of flops needed by the algorithm equals $14M_r-2$. It is worth noting that with an average Intel Core i750, the identification process takes 68.5 ms for $M_r=50000$; hence, the algorithm can be implemented in practice.
\section{Results\label{sec:Results}}
In this section, the results for simulated and off-the-air signals are presented.
\subsection{CCF for Simulated Signals}\label{sec:sim}
\begin{figure}
\centering
\includegraphics[width=1\linewidth]{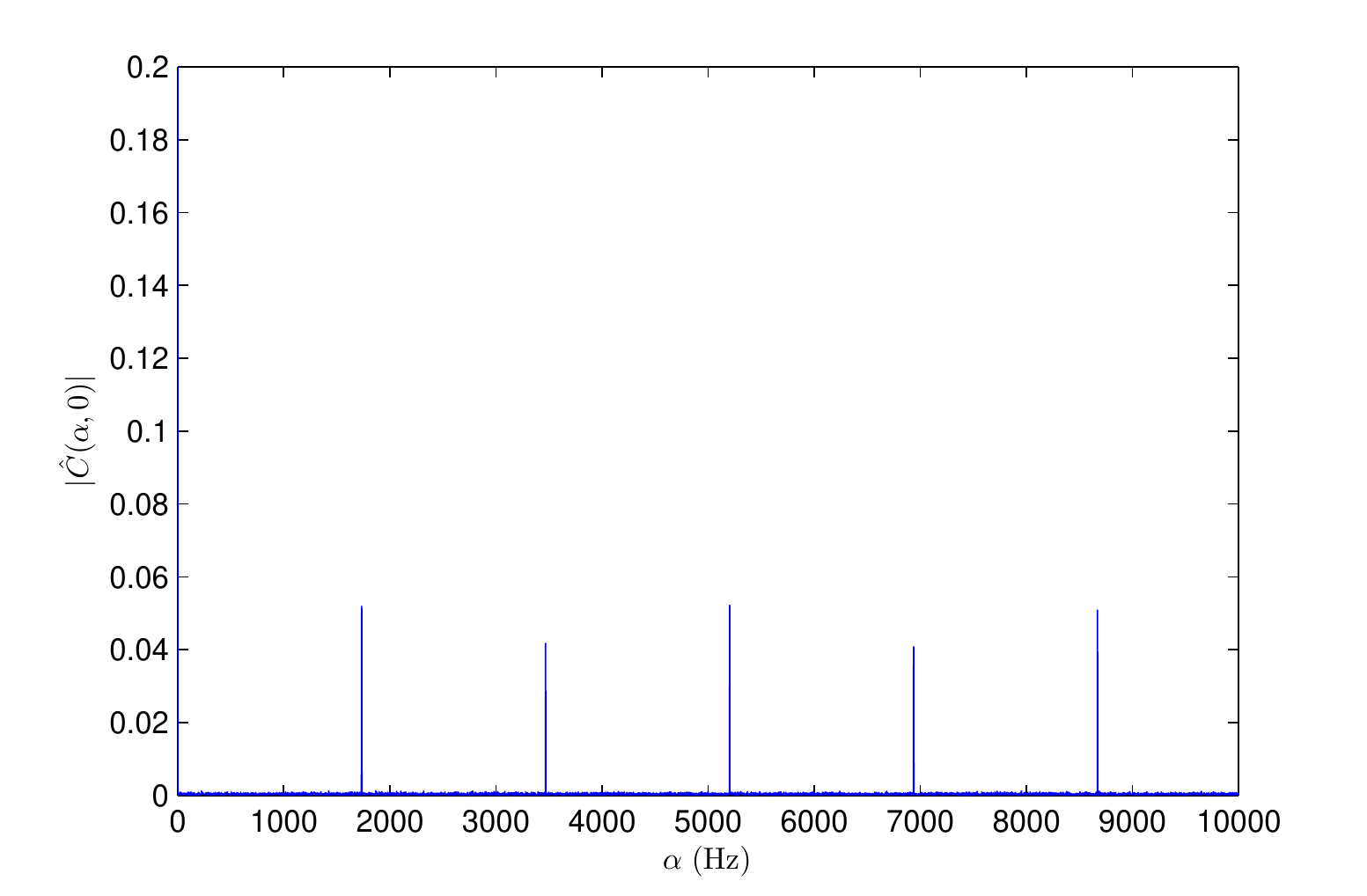}
\caption{CCF magnitude vs. CF for simulated GSM signals.}
\label{figGSM}
\vspace{-1.2em}
\end{figure}

\begin{figure}
\centering
\includegraphics[width=1\linewidth]{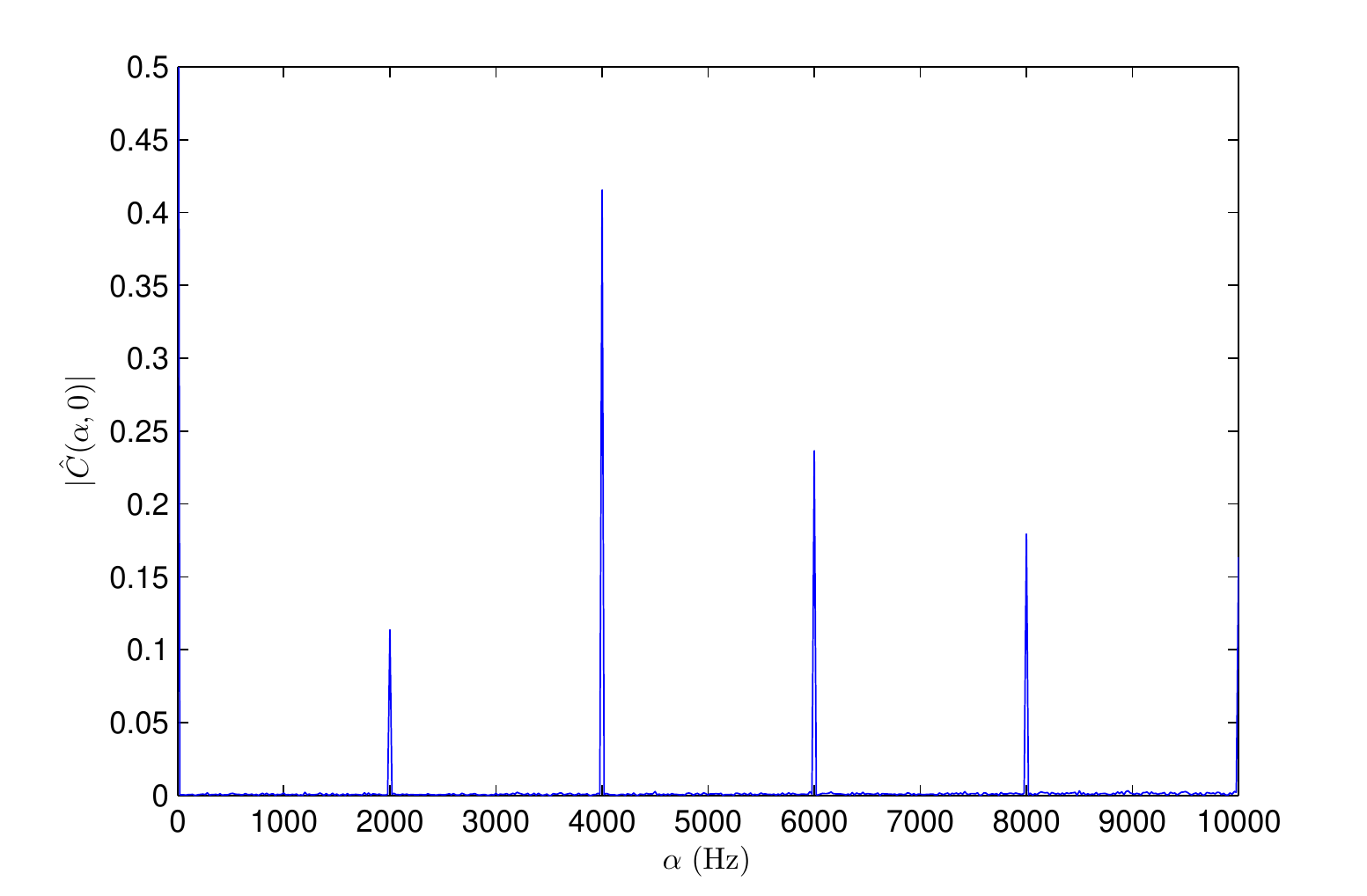}
\caption{CCF magnitude vs. CF for simulated LTE signals.}
\label{figLTE}
\vspace{-1.2em}
\end{figure}

Here we present simulation results for the CCF magnitude of the GSM and  LTE signals. For each case, a signal burst of 1000 time slots is generated and then transmitted through a frequency-selective fading channel consisting of $L_{p}=4$ statistically independent taps, each being a zero-mean complex Gaussian random variable. The channel is characterized by an exponential
power delay profile, $\sigma^{2}(p)=B_{h}exp(-p/5),$ where $p=0,...,L_p-1$
and $B_{h}$ is chosen such that the average power is normalized to unity and SNR is 20 dB. 
Fig. \ref{figGSM} presents simulation results for the GSM signals, while Fig. \ref{figLTE} shows results for the LTE signals. As expected, one can easily see that the CCF obtained from the simulated GSM signal has peaks at CFs equal to multiple integers of 1733 Hz, which is the reciprocal of the GSM time slot duration.
Furthermore, also as expected, the estimated CCF for the simulated LTE signal has peaks at CFs equal to multiple integers of 2 kHz, which is the reciprocal of the LTE time slot duration.
\subsection{CCF for Off-the-air Signals}
In this section, results for the CCF magnitude estimated from the signals received by a WSA4000 receiver is presented. The location of measurements was the ThinkRF company, in the north Kanata area of Ottawa, Canada. For each frequency band, $10^6$ samples were received. The bandwidth of the signal received by the WSA4000 receiver was 125 MHz, and the system had a decimation rate parameter to decrease the bandwidth; as such, depending on the expected bandwidth of the received signal, an appropriate decimation factor was considered. Please note that the proposed algorithm does not need to know the exact bandwidth of the received signal; as long as the signal of interest is in the bandwidth of the received signal, the proposed algorithm can identify it.\par
\begin{figure}
\centering
\includegraphics[width=1\linewidth]{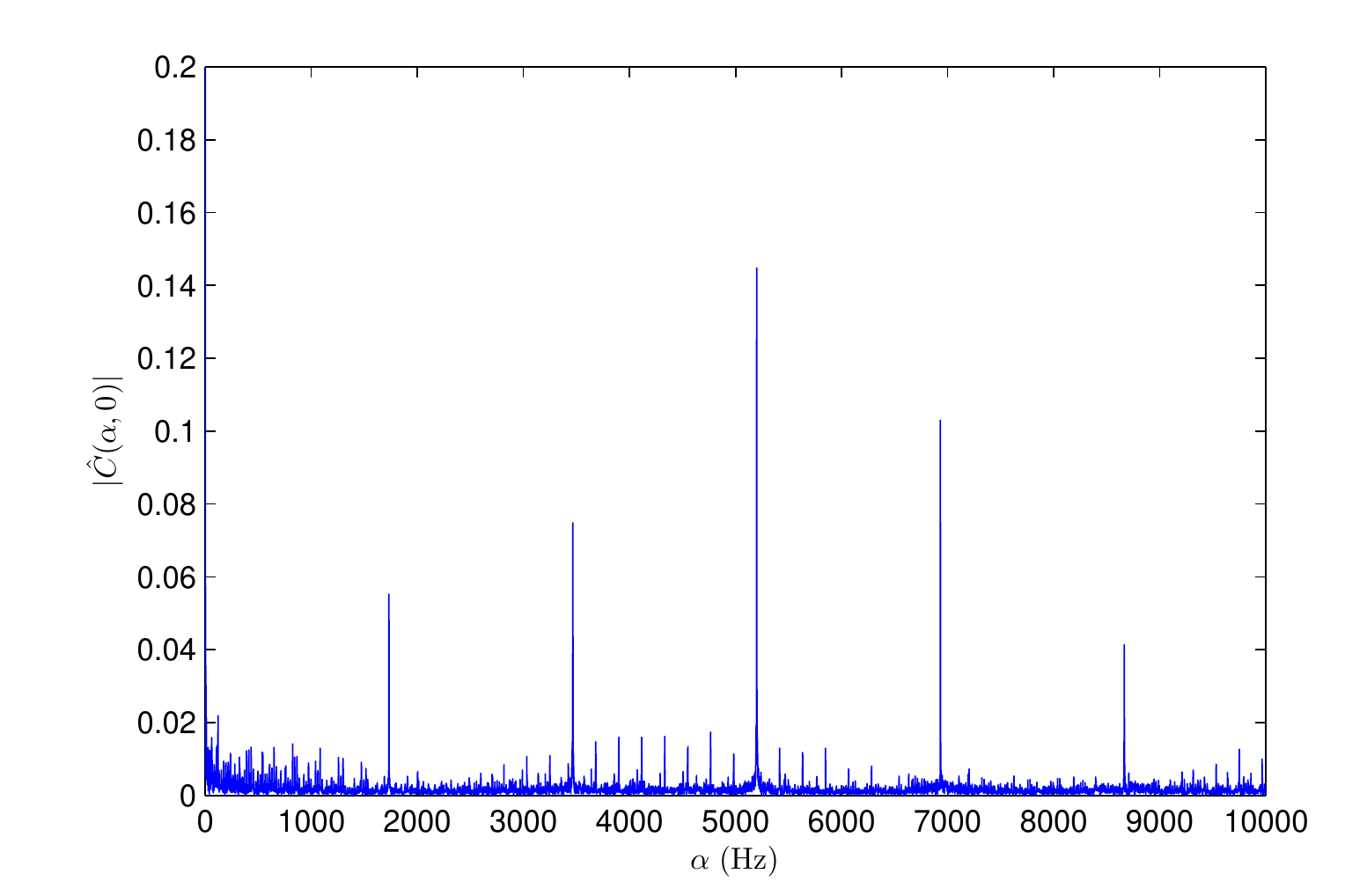}
\caption{CCF magnitude vs. CF for a signal received by the WSA4000 system within the frequency band of 869 MHz.}
\label{fig:869}
\vspace{-1.2em}
\end{figure}

Fig. \ref{fig:869} presents the CCF magnitude results for the signal in the 869 MHz band, where we expect the GSM signal from the Rogers base station (BS) located at approximately 460 meter away from our receiver. The decimation factor for this measurement was 64, corresponding to a 1.951 MHz receive bandwidth. This bandwidth was enough to cover the GSM bands supported by the corresponding BS$^1$\footnote{$^1$Each GSM band is 200 kHz.}. From Fig. \ref{fig:869}, one can see that the CCF estimated from the off-the-air GSM signal has peaks at CFs equal to multiple integers of 1733 Hz, which agrees with the simulation results.

\begin{figure}
\centering
\includegraphics[width=1\linewidth]{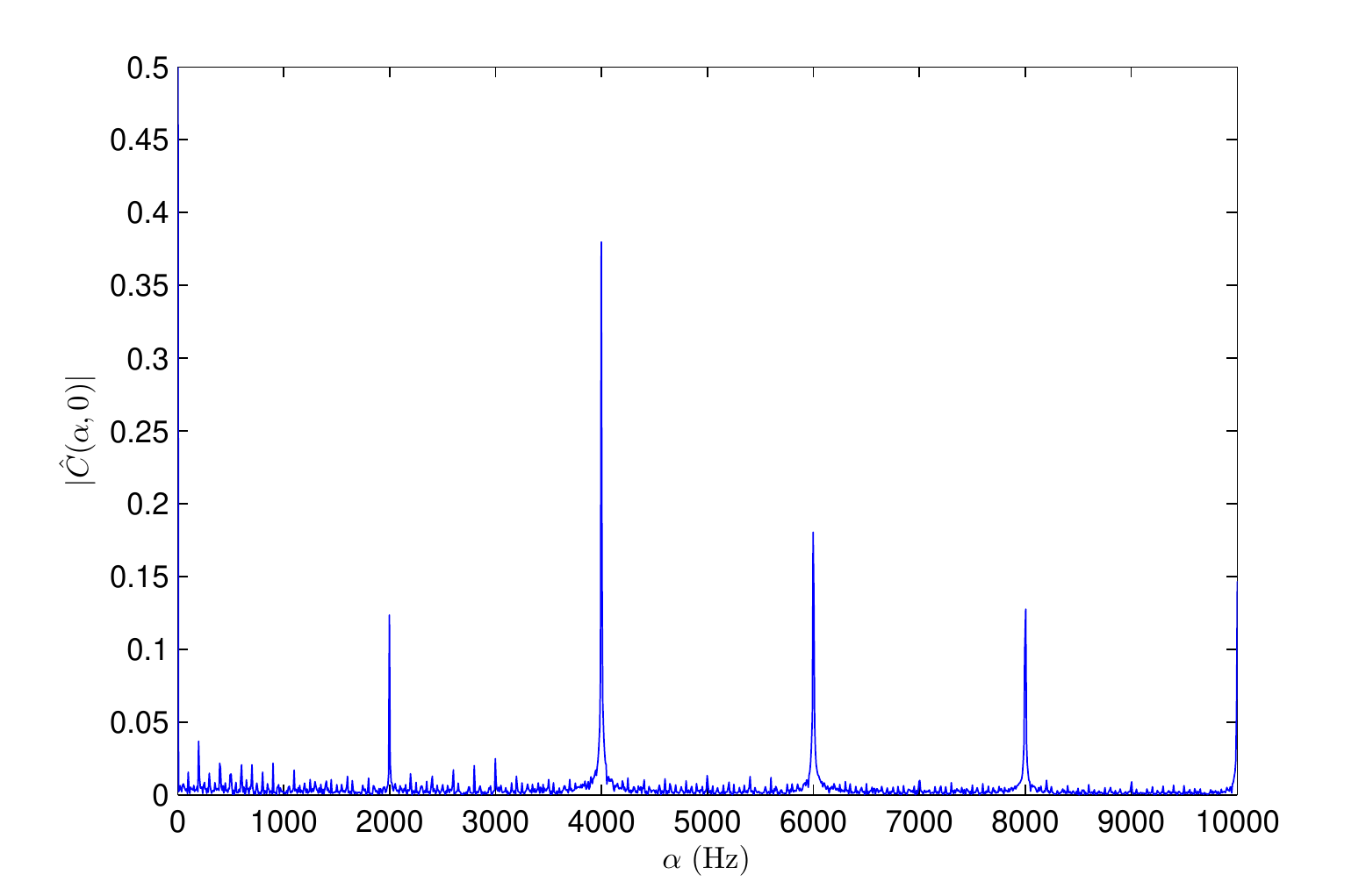}
\caption{CCF magnitude vs. CF for a signal received by the WSA4000 system within the frequency band of 2115 MHz.}
\label{fig:2115}
\vspace{-1.2em}
\end{figure}

Fig. \ref{fig:2115} presents the CCF magnitude results for the signal in the 2115 MHz band, where we expect the LTE signal from the Rogers BS located at approximately 460 meter away from the receiver (the location of this BS is the same as in the previous case). The decimation factor for this measurement was 16, corresponding to a 7.81 MHz receive bandwidth, which covers the LTE signal transmitted by the BS. From Fig. \ref{fig:2115}, one can see that the CCF estimated from the off-the-air LTE signal has peaks at CFs equal to multiple integers of 2 kHz, which concurs with the simulation results.

\subsection{Performance of the Proposed Algorithm}
In this section, the performance of the proposed algorithm for the identification of GSM and LTE signals is evaluated by Monte Carlo simulation through averaging over 1000 iterations. The simulation parameters are the same as in sub-section \ref{sec:sim}. The threshold is set up based on the constant false alarm criterion, and in each iteration, data is generated with a random timing offset taken from a uniform distribution within the first time slot.
\par
\begin{figure}
\centering
\includegraphics[width=1\linewidth]{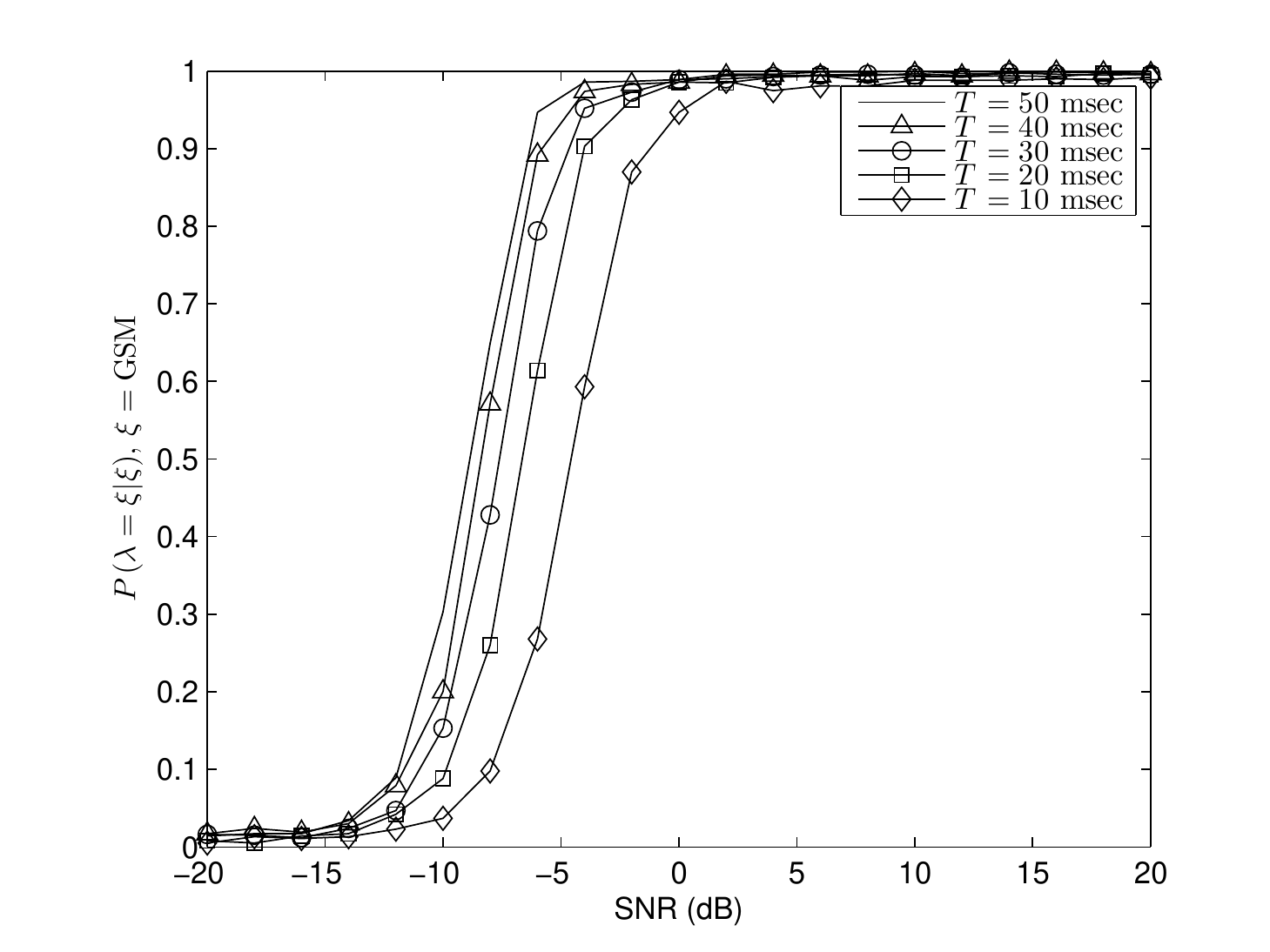}
\caption{Probability of correct identification for the GSM signals, $P(\lambda=\xi|\xi), \, \xi=\mbox{GSM}$, versus SNR for different observation times, $T$.}
\label{fig:GSM}
\vspace{-1.2em}
\end{figure}
\begin{figure}
\centering
\includegraphics[width=1\linewidth]{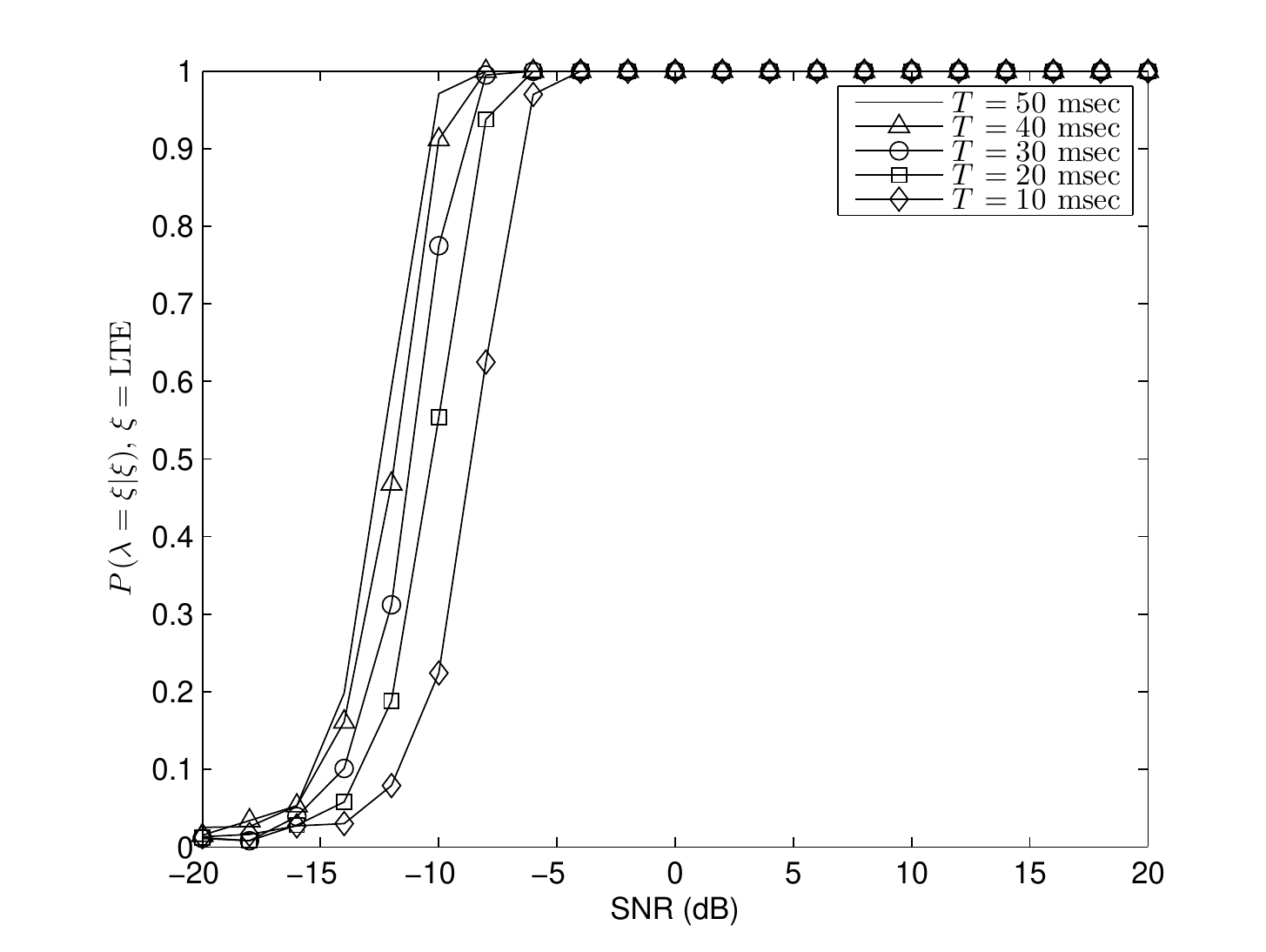}
\caption{Probability of correct identification for the LTE signals, $P(\lambda=\xi|\xi), \, \xi=\mbox{LTE}$, versus SNR for different observation times, $T$.}
\label{fig:LTE}
\end{figure}

Fig. \ref{fig:GSM} presents the performance of the proposed algorithm for the identification of the GSM signals for different observation times, with $P_F=10^{-2}$. From Fig. \ref{fig:GSM}, one can see that with SNR $>$ 0 dB, the probability of correct identification, $P(\lambda=\mbox{GSM}|\mbox{GSM})$, approaches one at an observation time as low as 10 ms, while with 50 ms, this occurs at about\linebreak-5 dB SNR.
Fig. \ref{fig:LTE} presents the performance of the proposed algorithm for the detection of the LTE signals for different observation times, with $P_F=10^{-2}$. From Fig. \ref{fig:LTE}, one can notice that with SNR $>$ -5 dB, the probability of correct correct identification, $P(\lambda=\mbox{LTE}|\mbox{LTE})$, approaches one at an observation time as low as 10 ms. In all cases, the results obtained for the LTE signal is better than for the GSM signal; however, one can obtain a good performance at short observation times and with low SNR for both signal types.

\begin{figure}
\centering
\includegraphics[width=1\linewidth]{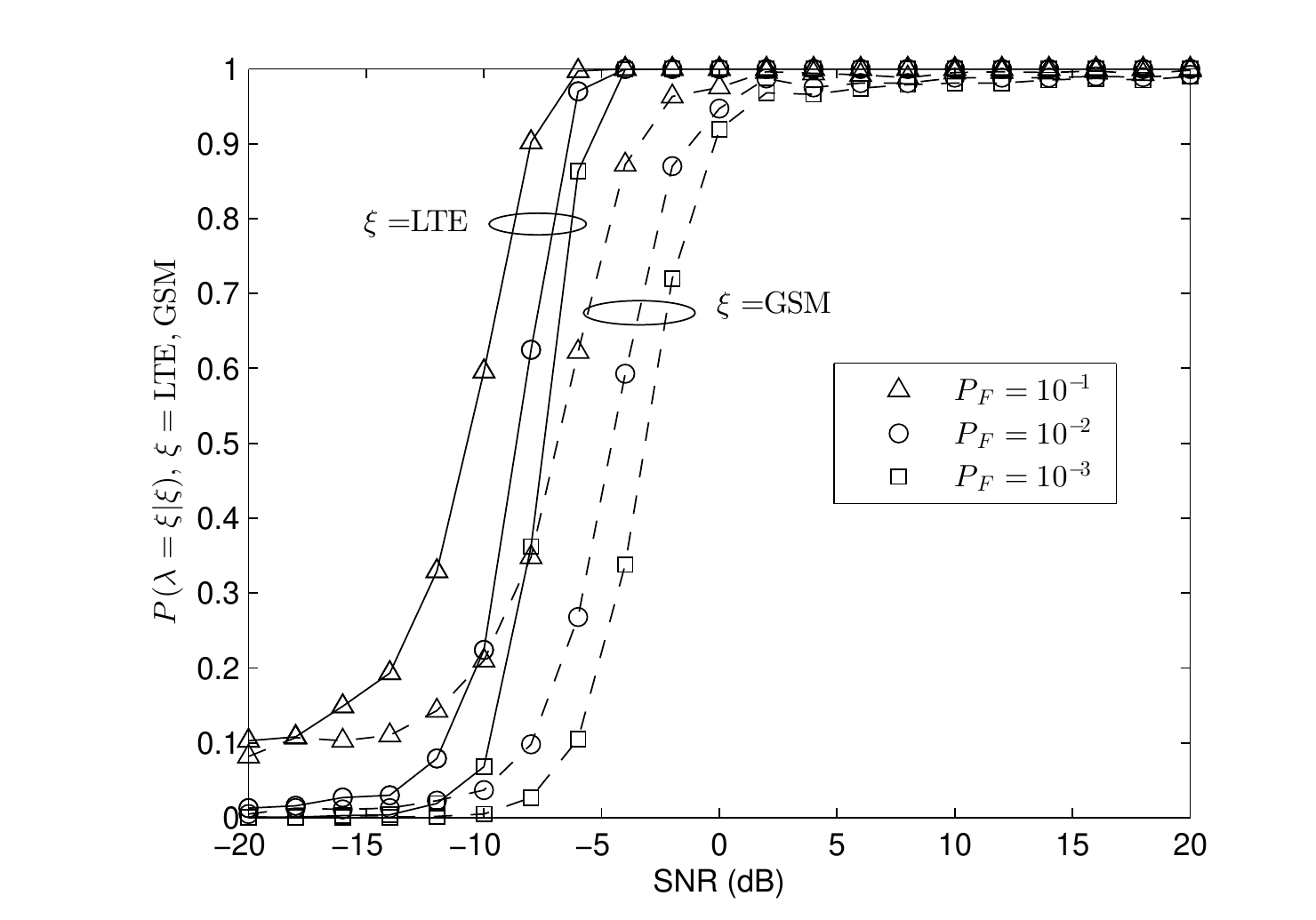}
\caption{Probability of correct identification for the GSM and LTE signals, $P(\lambda=\xi|\xi), \, \xi=\mbox{GSM}, \, \mbox{LTE}$, versus SNR for different $P_F$ values. Solid lines are used for the LTE signal and dashed lines are used for the GSM signal.}
\label{fig:PF}
\end{figure}

Fig. \ref{fig:PF} presents the performance of the proposed algorithm for the identification of the GSM and LTE signals for different $P_F$ values, with an observation time of $T=10$ ms. From Fig. \ref{fig:PF}, one can see that for LTE signals with SNR $>$ -5 dB, a very good performance is achieved regardless of the $P_F$ value; at lower SNRs, it is observed that $P(\lambda=\mbox{LTE}|\mbox{LTE})$ improves as $P_F$ increases. For the GSM signal, $P(\lambda=\mbox{GSM}|\mbox{GSM})$ approaches one for SNR $>$ 0 dB regardless of the $P_F$ value; at lower SNRs, the performance also enhances as $P_F$ increases. In all cases, a better performance is attained for LTE signal identification when compared with GSM.

\section{Conclusion\label{sec:Conclusion}}
In this paper, we proposed a very low complexity second-order cyclostationarity based algorithm for the identification of the GSM and LTE standard signals, which are commonly used in Canada. The proposed algorithm attains a very good performance at low  SNRs and with short observation times. Signals acquired by a ThinkRF WSA4000 receiver were used to prove the concept.
\section*{Acknowledgment}
The authors would like to acknowledge Tim Hember and Dr. Tarek Helaly from ThinkRF Corp. for their kind support.
\bibliographystyle{IEEEtran}

\end{document}